# Breaking the accuracy and resolution limitation of filter- and frequency-to-time mapping-based time and frequency acquisition methods by broadening the filter bandwidth

Pengcheng Zuo[a,b,†], Dong Ma[a,b,†], Xiaowei Li[a,b], and Yang Chen[a,b,*]

[a] Shanghai Key Laboratory of Multidimensional Information Processing, East China Normal University, Shanghai, 200241, China
[b] Engineering Center of SHMEC for Space Information and GNSS, East China Normal University, Shanghai, 200241, China
[*] ychen@ce.ecnu.edu.cn
[†] These authors contributed equally to this paper

**ABSTRACT**
In this paper, the filter- and frequency-to-time mapping (FTTM)-based photonics-assisted time and frequency acquisition methods are comprehensively analyzed and the accuracy and resolution limitation in the fast sweep scenario is broken by broadening the filter bandwidth. It is found that when the sweep speed is very fast, the width of the generated pulse via FTTM is mainly determined by the impulse response of the filter. In this case, appropriately increasing the filter bandwidth can significantly reduce the pulse width, so as to improve the measurement accuracy and resolution. FTTM-based short-time Fourier transform (STFT) and microwave frequency measurement using the stimulated Brillouin scattering (SBS) effect is demonstrated by comparing the results with and without SBS gain spectrum broadening and the improvement of measurement accuracy and frequency resolution is well confirmed. The frequency measurement accuracy of the system is improved by around 25 times compared with the former work using a similar sweep speed, while the frequency resolution of the STFT is also much improved compared with our former results.

**Keywords:** Time and frequency acquisition, frequency-to-time mapping, short-time Fourier transform, frequency measurement, filter bandwidth.

## 1. Introduction
Photonics-assisted microwave measurement applies photonics technology to the field of microwave measurement, which has the advantages of good tunability and reconfigurability, large bandwidth, and immunity to electromagnetic interference [1, 2]. Photonics-assisted one-dimensional frequency measurement is required to recognize stationary signals, such as single- or multi-frequency microwave signals. To realize microwave frequency measurement, the frequency-domain information needs to be mapped to parameters that are easier to measure. Frequency-to-space mapping [3-5], frequency-to-power mapping [6-8], and frequency-to-time mapping (FTTM) [9-21] are commonly used to achieve the above parameter mapping to obtain the frequency of the signal under test (SUT). Among them, FTTM-based methods have attracted great interests due to their intuitiveness, simplicity, and efficiency, which can be roughly divided into two categories in principle: dispersion-based FTTM [9-12] and filter-based FTTM [13-21]. The key devices in the first category are diverse dispersion devices, such as linearly chirped fiber Bragg grating and dispersion compensation fiber, which are used to separate different frequency components of the SUT in the time domain. It is particularly pointed out that the Fourier transform based on dispersion is also an FTTM process [11, 12]. The second category, i.e., the filter-based FTTM system, is more flexible and reconfigurable due to the good reconfigurability of microwave photonic filters and the frequency-sweep signal used to realize the FTTM, which can overcome the limitations of the measurement system caused by the difficulty of reconstructing the dispersive medium in the first category.

The FTTM-based one-dimensional frequency measurement provides the possibility to obtain the two-dimensional time-frequency information by expanding the dimensions of the measurement system. For example, starting from the one-dimensional Fourier transform system based on dispersion-based FTTM, two-dimensional short-time Fourier transform (STFT) [22-24] were realized and demonstrated. However, same as the dispersion-based FTTM in one-dimensional measurement, the two-dimensional STFT in [22-24] is also severely limited by the dispersion medium in reconfigurability. In addition, it is hard to achieve large operating bandwidth and good frequency resolution because of the limited dispersion value. Recently, to avoid the limitations of the existing dispersion-based STFT systems, we have chosen the filter-based FTTM as a starting point and demonstrated the first stimulated Brillouin scattering (SBS)-based STFT system that is free from dispersion [25]. A dynamic frequency resolution of 60 MHz was obtained when the frequency-sweep signal was swept at a sweep rate of 1 GHz/µs and an observation bandwidth of 12 GHz was achieved which was only limited by the equipment used in the experiment. Nevertheless, if the sweep rate of the frequency-sweep signal was increased, the frequency resolution would be degraded.

Measurement performance, such as frequency resolution and accuracy, is the key indicator of microwave measurement systems. The basic process of the filter- and FTTM-based measurement system is to map the frequency information of the SUT to the time domain in the form of low-speed electrical pulses, in which the desired frequency information can be obtained by locating the position of the pulses. During the FTTM process, the relative pulse width is generally considered to be a key factor affecting the frequency resolution and accuracy of the system and how to improve the frequency resolution and the accuracy of the system has always been concerned.  Filter in conjunction with frequency sweeping is widely employed for FTTM-based microwave frequency measurement. In [13], fiber Bragg grating pair composed Fabry–Perot (FP) filter and a tunable laser having a sweep rate of $1.25\times10^{-4}$ GHz/µs were used for microwave frequency measurement, in which the measurement accuracy and resolution were below 90 and 200 MHz, respectively. In [14], an integrated silicon photonic scanning filter having a 3-dB bandwidth of 325 MHz was swept at a sweep rate of $3.3\times10^{-3}$ GHz/µs for microwave frequency measurement, and the measurement resolution was 375 MHz and the error was 237.3 MHz. In [15], the sweep rate of a fiber

FP filter was set to 5.4×10$^{-3}$ GHz/μs, obtaining a measurement resolution is 90 MHz. To improve the measurement accuracy, recently, we have proposed an accuracy-improved multiple-frequency microwave measurement approach based on SBS filter- and FTTM-based structure [16], in which the measurement accuracy is greatly improved to better than ±1 MHz and the frequency resolution is around 20 MHz. To further improve the frequency resolution, we have demonstrated a multiple-frequency microwave measurement system with a frequency resolution of better than 10 MHz using a bandwidth-reduced SBS gain spectrum [17]. It should be pointed out that the sweep rate in [16, 17] is also slow, i.e., 5×10$^{-5}$ and 1.6×10$^{-5}$ GHz/μs, which also requires a relatively long time for frequency measurement. In this case, narrowing the bandwidth of the filter to a certain extent can improve the frequency resolution and accuracy of the system. Since the bandwidth of the optical filter, for example, the SBS filter and the FP filter, is difficult to be narrowed to less than several megahertz, an electrical filter with a much narrower bandwidth than that of a commonly used optical filter is employed and used in conjunction with frequency sweeping to further improve the frequency resolution and accuracy of the filter- and FTTM-based frequency measurement methods. For example, in [19], an electrical bandpass filter (BPF) with a narrow bandwidth of 750 kHz was used, and a resolution of better than 1 MHz and a measurement accuracy of ±0.4 MHz over a frequency range of 10 GHz were achieved under a sweep rate of 2.5×10$^{-6}$ GHz/μs. As discussed above, most of the previously reported filter- and FTTM-based photonics-assisted microwave measurement systems are demonstrated under a relatively low sweep rate of hundreds or tens of gigahertz per second or even less. Under such a slow sweep rate, good frequency resolution and accuracy can be achieved. However, in some applications that require high real-time measurement, such as electronic warfare systems, the sweep rate is a key indicator. In these applications, a slow sweep rate is unacceptable, and there is the possibility of missing detection of burst signals and non-stationary signals. Few works performed fast sweep measurements to meet the requirements of real-time measurement. For example, in [20], the sweep rate was around 0.18 GHz/μs, the measurement resolution is 60 MHz, and the accuracy is better than ±60 MHz. In [21], the sweep rate of an FDML laser was 2.74 GHz/μs, and a measurement resolution of 200 MHz and an accuracy of better than ±100 MHz were achieved. It can be seen that when the sweep rate is high, the measurement accuracy and resolution are much worse than the best results achieved when the sweep rate is low as demonstrated in [16, 17]. Besides the one-dimensional frequency measurement, the filter- and FTTM-based two-dimensional STFT proposed by us in [25] also requires a high sweep rate to guarantee that the signal in each sweep period can be approximated as a stationary signal. Thus, in [25], a high sweep rate of 4 GHz/μs was employed and the frequency resolution was also greatly limited by the high sweep rate.

Therefore, new methods are urgently needed to solve the problem of poor measurement resolution and accuracy when the sweep rate is high both in one-dimensional frequency measurement and two-dimensional time and frequency measurement. In this paper, for the first time, the accuracy and resolution limitation of filter- and FTTM-based time and frequency acquisition methods in the fast sweep scenario is broken by broadening the filter bandwidth. The principle of FTTM realized by scanning a filter using a frequency-sweep signal is further analyzed. We find that when the sweep speed is very fast, the width of the generated pulse via FTTM is mainly determined by the impulse response of the filter. In this case, appropriately increasing the filter bandwidth can significantly reduce the pulse width, to improve the measurement accuracy and resolution. FTTM-based STFT and frequency measurement using the SBS effect is demonstrated by comparing the results with and without SBS gain spectrum broadening and the improvement of measurement accuracy and frequency resolution is well confirmed. The frequency measurement accuracy of the system is improved by around 25 times compared with the works using similar sweep speed [21], while the frequency resolution of the STFT is also doubled or tripled compared with our former work [25].

## 2. Principle

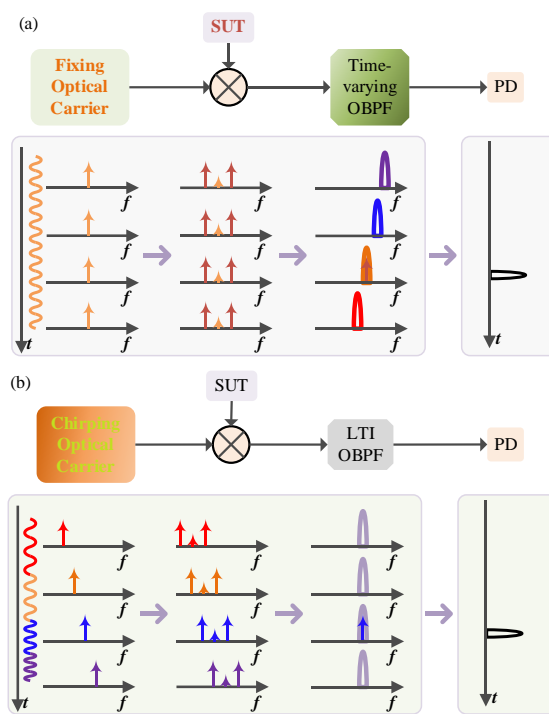

Fig. 1. Schematics of two kinds of the filter- and FTTM-based time and frequency acquisition system. SUT, signal under test; PD, photodetector; LTI, linear time-invariant; OBPF, optical bandpass filter.

In the filter- and FTTM-based time and frequency acquisition system, the key part is to map different frequency components to low-frequency electrical pulses at different times. As shown in Fig. 1, this process can be commonly implemented by two methods: One is to capture different frequencies to be measured carried by a fixed optical carrier at different times through a frequency-sweep filter as shown in Fig. 1(a); the other is to sweep the optical frequency of the signal to be measured through a chirped optical carrier and capture different frequencies using a fixed filter at different times as shown in Fig. 1(b).

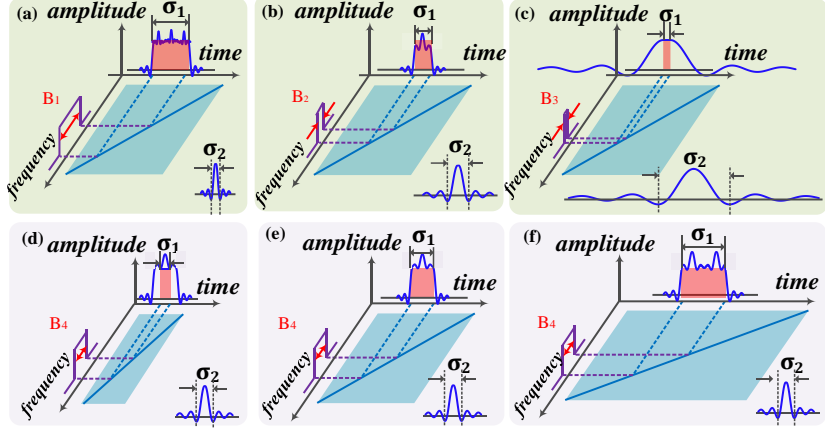

Fig. 2. Schematic diagram of the evolution of the generated pulse signal for frequency identification when frequency-sweep signals sweep through filters in different cases. (a)-(c) Fixed sweep rate and different filter bandwidths; (d)-(f) different sweep rates and fixed filter bandwidth.

The process of generating pulses through FTTM is further analyzed and shown in Fig. 2. Here, an ideal bandpass filter (BPF) is used to simplify the analysis process. As shown in Fig. 2, the impulse response of an ideal BPF is a sinc function, and the time width of the impulse response of the ideal BPF is inversely proportional to the filter bandwidth. When a frequency-sweep signal passes through the ideal BPF, it can be roughly considered that the interaction consists of two parts for ease of analysis: 1) Only the frequency components falling into the passband of the ideal BPF can pass through the filter and can be converted to low-frequency pulses; 2) because the impulse response of the ideal BPF is not a delta function, it will broaden the generated pulse. To simplify the analysis, the passing width representing how long the SUT passes through the BPF is denoted as

$$\sigma_1 = B/k, \tag{1}$$

and a broadening width of the BPF is defined as

$$\sigma_2 = 1/B. \tag{2}$$

The width of the generated pulses can be analyzed in the following cases: 1) Case 1: $\sigma_1 \ll \sigma_2$, the width of the generated pulse is mainly determined by the broadening width $\sigma_2$; 2) Case 2: $\sigma_2 \ll \sigma_1$, the width of the generated pulse is mainly determined by the passing width $\sigma_1$; 3) Case 3: $\sigma_1$ and $\sigma_2$ have relatively close values and the width of the generated pulse is jointly determined by both these two factors.

Figure 2(a) to (c) show the schematic of Case 2, Case 3, and Case 1, respectively. When the sweep rate is fixed, the greater the ideal BPF bandwidth, the wider the passing width $\sigma_1$. Nevertheless, with the decrease of the BPF bandwidth, as can be seen, the width of the impulse response (also the broadening width $\sigma_2$) increases. Thus, there is a limit to the reduction of the generated pulse width by reducing the filter bandwidth. When the bandwidth is reduced from $B_1$ to $B_2$ as shown in Fig. 2(a) and (b), the pulse width reduction caused by the decrease of $\sigma_1$ exceeds the pulse width increase caused by the increase of $\sigma_2$. Therefore, the overall pulse width is reduced in this case. However, as shown in Fig. 2(b) and (c), when the bandwidth is reduced from B2 to B3, although $\sigma_1$ is largely decreased, the increase of $\sigma_2$ is greater, leading to an overall increase in the pulse width even if the BPF bandwidth is decreased. In addition, when the sweep rate is fixed, the frequency represented by a fixed time interval is also fixed in the FTTM process. Therefore, in this case, the pulse width will directly determine the frequency resolution and accuracy. Same as the pulse width, the frequency resolution and accuracy also have a limit when a proper BPF bandwidth is selected and this optimal bandwidth is highly related to the chirp rate of the frequency-sweep signal.

Another case is then analyzed as shown in Fig. 2(d) to (f), in which the BPF bandwidth is fixed and the sweep rate is changed. Because the filter bandwidth is fixed, the broadening width $\sigma_2$ is the same in these three cases. However, due to the different chirp rates, the passing width $\sigma_1$ is also different. As shown in Fig. 2(d) and (e), when the chirp rate decreases, the passing width $\sigma_1$ increases in inverse proportion while the broadening width $\sigma_2$ is unchanged. If only the passing width $\sigma_1$ is considered as the pulse width, the frequency resolution and accuracy will be the same under different sweep rates because the frequency represented by a fixed time interval also changes inversely in the FTTM process under the same filter bandwidth. However, when the broadening width $\sigma_2$ is taken into account, the situation is completely different. Because the filter bandwidth is fixed, the broadening width $\sigma_2$ is a constant for different chirp rates. Therefore, the broadening width $\sigma_2$ has a larger influence on the total pulse width when the sweep rate is larger. As the sweep rate decreases, its influence is smaller and the increase of frequency resolution and accuracy will be decreased and approach a limitation. In the following of this work, the frequency resolution of the system is defined as the frequency value represented by the full width at half maximum (FWHM) of the generated pulse unless otherwise specified.

Since the total pulse width is jointly determined by $\sigma_1$ and $\sigma_2$, and $\sigma_1$ is proportional to and $\sigma_2$ is inversely proportional to the filter bandwidth, it is easy to know whether the sweep rate is as low as hundreds of Hertz per microsecond or as high as a few gigahertz per microsecond, the best frequency resolution can be achieved using a proper filter bandwidth, and naturally, a corresponding best measurement accuracy will be obtained. Moreover, the slower the sweep rate, the smaller the filter bandwidth required to achieve the best frequency resolution. In previous works, the sweep rates are relatively low. For example, when the sweep rate changes from $10^{11}$ Hz/s to $10^{12}$ Hz/s and from $10^8$ Hz/s to $10^9$ Hz/s, the best filter bandwidth is around several megahertz and tens of kilohertz. In these cases, except for the extremely low filter bandwidth, the frequency resolution increases with the decrease of filter bandwidth. When the filter bandwidth is smaller than the best values (megahertz or tens of kilohertz), the resolution can be improved by increasing the filter bandwidth until reaching the best filter bandwidth. Nevertheless, this is not found and used in previous works to increase the resolution because under a low sweep rate, such low filter bandwidth is hard to be realized, especially in the optical domain. In comparison, as the sweep rate increases, the best filter bandwidth that can achieve the best resolution is greatly increased. For example, when the sweep rate is from $10^{15}$ Hz/s to $10^{16}$ Hz/s, the best filter bandwidth is roughly around several tens of megahertz, which means further increasing the filter bandwidth from smaller values (several megahertz to ten or twenty megahertz) to these values can improve the frequency resolution of the measurement, as well as the measurement accuracy. Extending the bandwidth of the filter in such a range is easier to achieve. This is also the key point in this work for breaking the accuracy and resolution limitation of filter- and FTTM-based time and frequency acquisition methods.

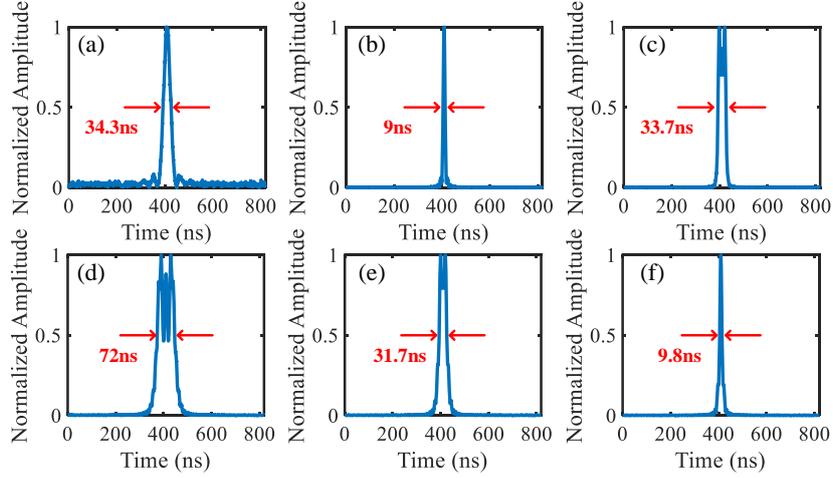

Fig. 3. Stimulation results of the generated electrical pulses under different sweep rates and different ideal BPF bandwidth. Fixed sweep rate of 10 GHz/μs and different filter bandwidths of (a) 25 MHz, (b) 100 MHz, and (c) 400 MHz. Fixed filter bandwidth of 100 MHz and different sweep rates of (d) 1 GHz/μs, (e) 2 GHz/μs, and (f) 4 GHz/μs.

Then, a physical simulation is conducted to show the waveform changes with the sweep rate and filter bandwidth. First, a linearly frequency-modulated (LFM) signal with a chirp rate of 10 GHz/μs is sent to an ideal BPF with a bandwidth of 25 MHz. According to Eqs. (1) and (2), the passing width $\sigma_1$ is 2.5 ns while the broadening width $\sigma_2$ is 40 ns, so the total width of the pulse is mainly determined by the broadening width. The FWHM of the pulse is measured to be around 34.3 ns from Fig. 3(a). When the filter bandwidth is increased to 100 MHz, the passing width $\sigma_1$ and the broadening width $\sigma_2$ are both 10 ns, so the FWHM of the pulse is jointly determined by $\sigma_1$ and $\sigma_2$, which is measured to be 9 ns from Fig. 3(b). If the bandwidth of the BPF is further increased, for example, to 400 MHz, the passing width $\sigma_1$ is 40 ns while the broadening width $\sigma_2$ is changed to 2.5 ns, resulting in that the total pulse width is mainly determined by the passing width $\sigma_1$. As shown in Fig. 3(c), the FWHM of the pulse is around 33.7 ns. The above simulation results are consistent with the analysis in Fig. 2 (a) to (c). Figure 3(d) to (f) show the generated pulses when the ideal BPF bandwidth is fixed at 100 MHz and the sweep rate is 1, 2, and 4 GHz/μs.

In the above analysis, the ideal BPF is employed for simplicity. It is concluded that, when the sweep rate is fixed, a proper bandwidth can be found to generate the narrowest pulse for the best measurement resolution and accuracy. We reveal that when the frequency-sweep signal is swept at very high speeds, optimum measurement performance can be obtained by increasing the width of the optical or electrical filters, which will be demonstrated in the following part of the paper using the microwave photonic filter constructed by the SBS effect.

## 3. Experiment results and discussion
*3.1 Experimental setup*

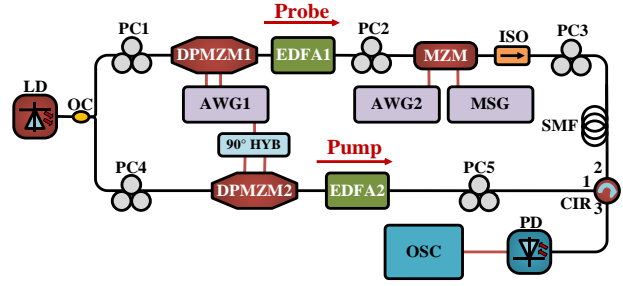

Fig. 4. Schematic diagram of the proposed SBS-based time and frequency acquisition system. LD, laser diode; OC, optical coupler; PC, polarization controller; DP-MZM, dual-parallel Mach–Zehnder modulator; MZM, Mach–Zehnder modulator; AWG, arbitrary waveform generator; MSG, microwave signal generator; EDFA, erbium-doped fiber amplifier; ISO, isolator; SMF, single-mode fiber; PD, photodetector; OSC, oscilloscope.

To verify the proposed approach, proof-of-concept experiments based on the setup shown in Fig. 4 for implementing the SBS-based time and frequency acquisition system are performed. A 15.5-dBm optical carrier centered at 1551.303 nm from a laser diode (LD, ID Photonics CoBriteDX1-1-C-H01-FA) is divided into two paths via a 50/50 optical coupler (OC). In the upper path, the optical carrier is carrier-suppressed single-sideband modulated at a dual-parallel Mach–Zehnder modulator (DP-MZM1, Fujitsu FTM 7961EX) by a sweep electrical signal from an arbitrary waveform generator (AWG1, Keysight M8195A) to generate a sweep optical signal. The output of DP-MZM1 is injected into a null-biased Mach–Zehnder modulator (MZM, Fujitsu FTM 7938) after being amplified by an erbium-doped fiber amplifier (EDFA1, MAX-RAY EDFA-PA-35-B), and then carrier-suppressed double-sideband modulated by a combined signal from the electrical coupler (EC, Narda 4456-2), which includes the SUT from AWG2 (Keysight M8190A) and a fixed reference from the microwave signal generator (MSG, Agilent 83630B). Then, the output of the MZM is injected into the 25.2-km single-mode fiber (SMF) through the isolator. In the lower path, the CW light wave from the LD is used as the pump wave. To manipulate the SBS gain bandwidth, DP-MZM2 (Fujitsu, FTM 7961EX) is inserted into the pump path and modulated by an electrical frequency-sweep signal from AWG1 with different sweep ranges, and then reversely launched into the 25.2-km SMF after being amplified by EDFA2 (Amonics, EDFA-PA-35-B), where it interacts with the counter-propagating probe wave from the upper branch. PC3 and PC5 are used to ensure the most efficient stimulated Brillouin interaction. Then, the optical signal from the SMF is detected by a photodetector (PD, Nortel PP-10G) and monitored by an oscilloscope (OSC, R&S RTO2032).

*3.2 Width of the time-domain pulses*
First, the broadening of the SBS gain bandwidth is demonstrated. To broaden the SBS gain bandwidth, in Fig. 4, DP-MZM2 is modulated by an electrical frequency-sweep signal. Here, the electrical frequency-sweep signal is centered at 6 GHz and with different sweep ranges of 0, 30, 60, 90, 120, 150, 180, and 210 MHz. The SBS gain spectrum is measured by mapping it to a microwave photonic bandpass filter via phase-to-intensity-modulation conversion. The response is measured using a vector network analyzer (VNA, Agilent 8720ES) and shown in Fig. 5. As can be seen from Fig. 5(a), when the sweep range is changed from 0 to 210 MHz, the SBS gain spectrum is broadened from the natural Lorentz-shaped spectrum to a wideband spectrum with a flat top. The increase of the 3-dB gain bandwidths is measured and shown in Fig. 5(b). In the following part, the same method is employed to increase the SBS gain bandwidth for measurement performance improvement.

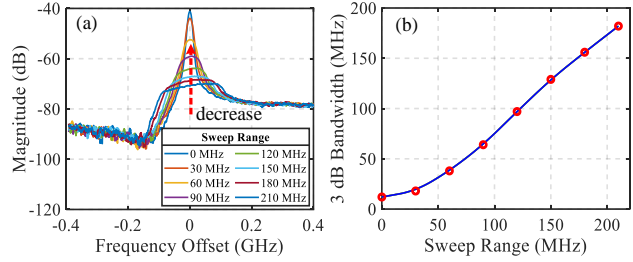

Fig. 5. (a) Measured frequency response of the SBS-based microwave photonic bandpass filter when the frequency-sweep pump wave has different sweep ranges of 0, 30, 60, 90, 120, 150, 180, and 210 MHz. (b) The corresponding 3-dB SBS gain bandwidth.

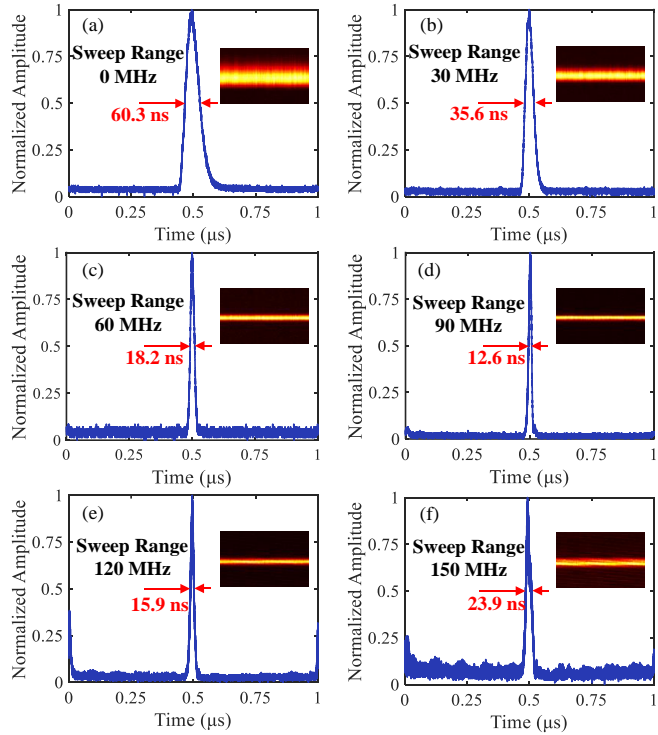

Fig. 6. Measured time-domain pulses and corresponding time-frequency diagrams of a single-tone signal by the SBS filter- and FTTM-based system configured with an electrical frequency-sweep signal with a bandwidth ranging from 4.8 to 8.8 GHz and 1-μs period in the probe path. The sweep range of the pump wave is (a) 0, (b) 30, (c) 60, (d) 90, (e) 120, and (f) 150 MHz.

The width of the time-domain pulses after SBS-based FTTM is studied when the 3-dB SBS gain bandwidth is broadened. First, the sweep rate is set to 4 GHz/μs as an example. The system is configured with an electrical frequency-sweep signal in the probe path with a bandwidth ranging from 4.8 to 8.8 GHz and a 1-μs period. Single-tone test is performed to observe the evolution of the pulse width when the sweep range of the pump wave is 0, 30, 60, 90, 120, and 150 MHz, with the results shown in Fig. 6. As can be seen, with the increase of the SBS gain spectrum (sweep range of the pump wave from 0 to 90 MHz), the width of the time-domain pulse is narrowed from 60.3 to 12.6 ns, as shown in Fig. 6(a) to (d). Nevertheless, further increasing the sweep range of the pump wave from 90 to 150 MHz, as shown in Fig. 6(e) and (f), will broaden the pulse width from 12.6 to 23.9 ns. The time-frequency diagram of the single-tone signal after STFT analysis also shows the same trend, which is consistent with the analysis and simulation in Section 2.

To better confirm the effect of the broadened SBS gain bandwidth on the pulse width, and to further study the performance of broadening the SBS gain bandwidth at different sweep rates, the system is demonstrated under different pump wave sweep rates. Fig. 7(a) and (c) show the measured pulse width under different sweep rates of 1, 2, 4, 6, 8, and 10 GHz/μs in the probe path and different SBS gain bandwidths broadened by different pump wave sweep ranges. Fig. 7(b) and (d) show the corresponding frequency resolution by mapping the FWHM of the pulse to the frequency value it represents. With the change of the sweep range of the pump wave, i.e., the change of the SBS gain bandwidth, the narrowest pulse width, as well as the best frequency resolution, can be obtained at different SBS gain bandwidth when the sweep rate of the probe path is changed. Moreover, the higher the sweep rate of the probe path, the larger the SBS bandwidth required to obtain the narrowest pulse and best frequency resolution, which is consistent with our analysis in Section 2. It can also be inferred that, as given in our previous discussion, when the sweep rate in the probe path is slow enough, the SBS bandwidth to obtain the narrowest pulse width or the best resolution will also be small enough.

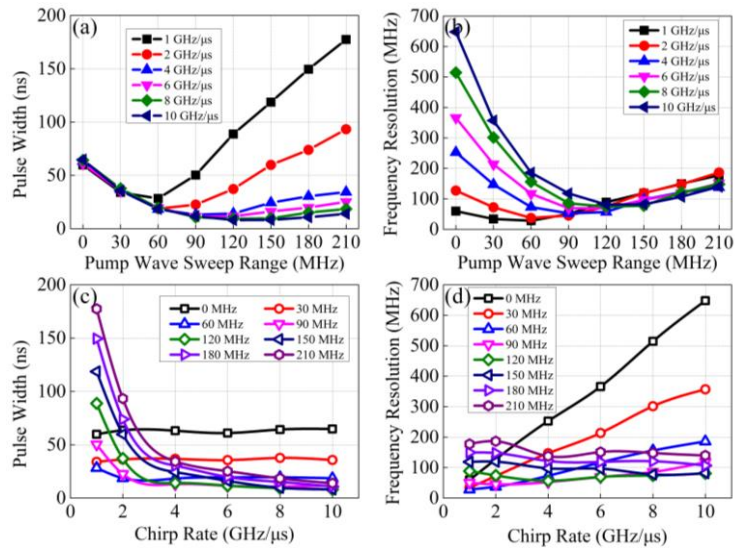

Fig. 7. Measured pulse width and corresponding frequency resolution when the sweep range of the pump wave and the sweep rate in the probe path are changed.

*3.3 SBS-based STFT verification*

SBS-based STFT is performed to verify the validity of the proposed approach. For implementing the SBS-based STFT, the system shown in Fig. 4 is configured with electrical linearly frequency-sweep signals. First, the sweep rate in the probe path is set to 1 GHz/μs by applying an electrical frequency-sweep signal with a bandwidth ranging from 4.8 to 5.8 GHz and a 1-μs period. The SUT is an LFM signal with a bandwidth ranging from 0 to 4 GHz and a time duration of 200 μs. It is obvious that the SUT cannot be completely measured in this case, because the bandwidth of the frequency-sweep signal is smaller than that of the SUT and the measurement range is from 0 to 1 GHz. As can be seen from Fig. 8(a) to (c), the frequency resolution becomes better as the pump wave sweep range is increased from 0 to 60 MHz, which indicates that broadening the SBS gain can improve the frequency resolution. Then, the system is configured with 1-μs period electrical frequency-sweep signals with bandwidths ranging from 4.8 to 6.8 GHz and 4.8 to 8.8 GHz, which means the sweep rates in the probe path are 2 and 4 GHz/μs and the corresponding measurement bandwidths are 2 and 4 GHz, respectively. When the sweep rate in the probe wave is 4 GHz/μs, the SUT can be measured completely. It is also observed that with the increase of the pump wave sweep range, the frequency resolution is also improved no matter whether the sweep rate in the probe path is 2 or 4 GHz/μs. Comparing the results with the same SBS gain bandwidth and different sweep rate of the probe path, it can be seen that when the SBS gain bandwidth is fixed by a pump wave sweep range of 0, 30, or 60 MHz, the frequency resolution become worse as the sweep rate in the probe path is increased from 1 to 4 GHz/μs. This is consistent with our analysis in Section 2 and the results given in Fig. 7.

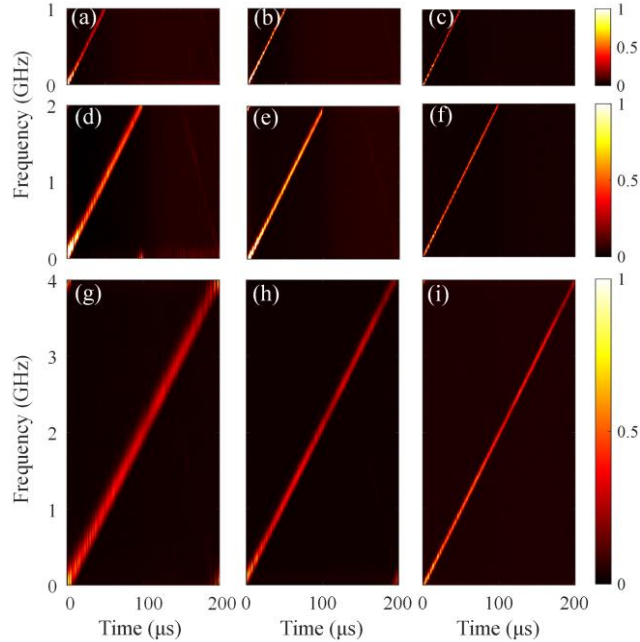

Fig. 8. Measured time-frequency diagrams by the SBS-based STFT system for an LFM signal with a bandwidth ranging from 0 to 4 GHz and a time duration of 200 μs. (a) (b) (c) The sweep rate in the probe path is 1 GHz/μs and the sweep ranges of the pump wave are 0, 30, and 60 MHz, respectively. (d) (e) (f) The sweep rate in the probe path is 2 GHz/μs and the sweep ranges of the pump wave are 0, 30, and 60 MHz, respectively. (g) (h) (i) The sweep rate in the probe path is 4 GHz/μs and the sweep range of the pump wave is 0, 30, and 60 MHz, respectively.

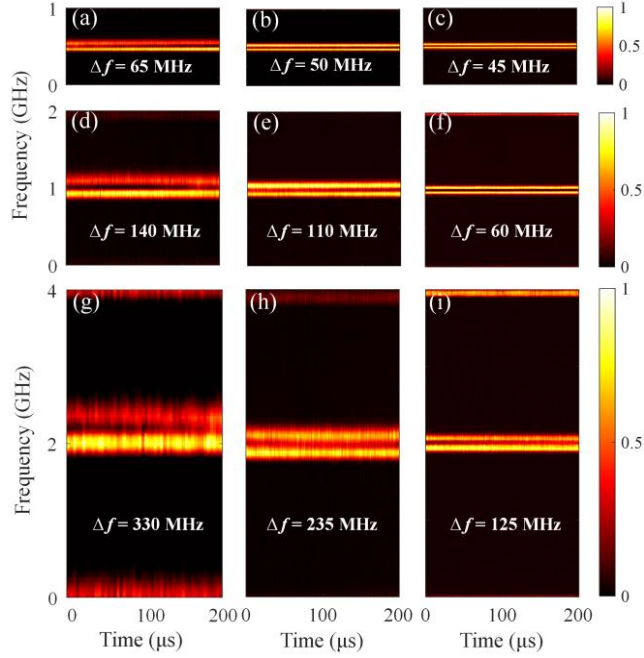

Fig. 9. Measured time-frequency diagrams by the SBS-based STFT for a two-tone signal with different frequency intervals. (a) (b) (c) The sweep rate in the probe path is 1 GHz/μs and the sweep ranges of the pump wave are 0, 30, and 60 MHz, respectively. (d) (e) (f) The sweep rate in the probe path is 2 GHz/μs and the sweep range of the pump wave is 0, 30, and 60 MHz, respectively. (g) (h) (i) The sweep rate in the probe path is 4 GHz/μs and the sweep range of the pump wave is 0, 30, and 60 MHz, respectively.

In the above analysis, the frequency resolution is defined according to the FWHM of the generated pulses. In [25], the frequency resolution was measured by employing a two-tone test. Then, the frequency resolution of the system is also demonstrated using two-tone signals with different frequency intervals under different conditions. In the experiment, the sweep rates of the frequency-sweep signals are set to 1, 2, and 4 GHz/μs, and the sweep ranges of the frequency-sweep pump wave are set to 0, 30, and 60 MHz, respectively. As shown in Fig. 9(a), to (c), when the sweep rate in the probe path is 1 GHz/μs and the sweep ranges of the pump wave is changed from 0 to 30 and 60 MHz, the frequency resolutions are better than 65, 50, and 45 MHz, respectively. These values change to 140, 110, and 60 MHz when the sweep rate is 2 GHz/μs and to 330, 235, and 125 MHz when the sweep rate is 4 GHz/μs. For all cases, the frequency resolution becomes better as the SBS gain bandwidth is increasing. Compared with the results in [25], the frequency resolution in this work is much improved thanks to the broadening of the filter bandwidth.

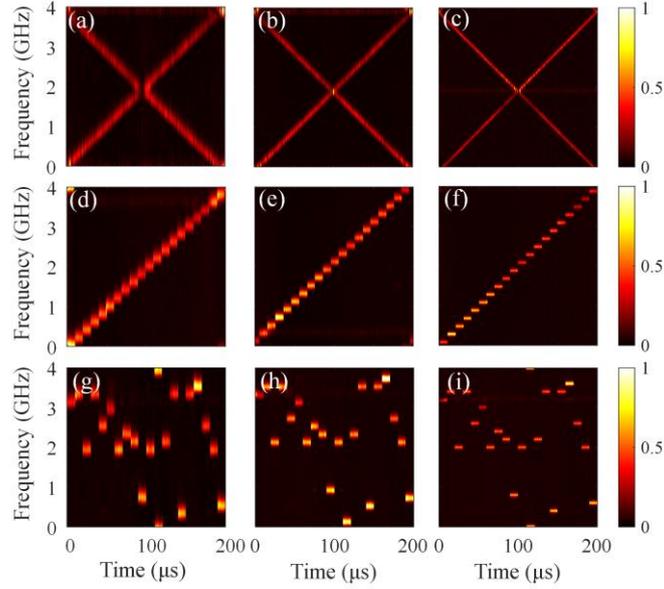

Fig. 10. Measured time-frequency diagrams by the SBS-based STFT for the dual-chirp LFM signal, step-frequency signal, and frequency-hopping signal. The sweep rate in the probe path is 4 GHz/μs. The sweep ranges of the pump wave are (a) (d) (g) 0, (b) (e) (h) 30, and (c) (f) (i) 60 MHz.

Time-frequency analysis of multi-format signals is demonstrated under different SBS gain bandwidth to show the frequency resolution improvement. In this experiment, the sweep rate in the probe path is fixed to 4 GHz/μs. Different kinds of broadband signals, including the dual-chirp LFM signal, the step-frequency signal, and the frequency-hopping signal, are chosen as the SUTs. As can be seen from Fig. 10, the time-frequency diagrams of all these signals are well constructed. More importantly, with the increase of the SBS gain bandwidth, the frequency resolutions of all these kinds of signals are greatly improved, which also confirms that the SBS-based STFT system is effective for time-frequency analyzing multi-format signals with a reconfigurable resolution by manipulating the bandwidth of the SBS gain spectrum.

*3.4 SBS-based frequency measurement verification*

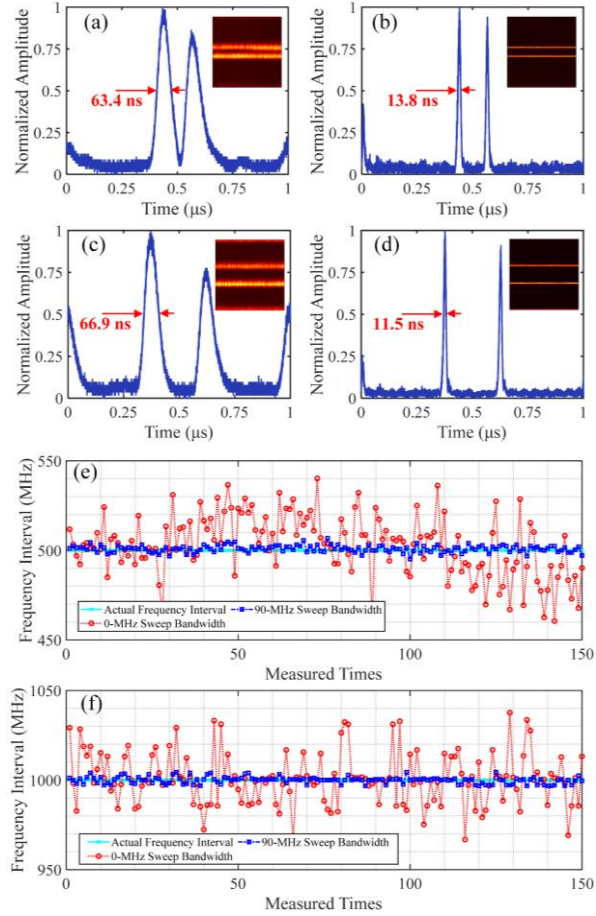

Fig. 11. Measured time-domain pulses and corresponding time-frequency diagrams of two microwave signals with a frequency interval of (a) (b)500 MHz and (c) (d)1 GHz. The sweep ranges of the pump wave are (a) (c) 0 MHz and (b) (d) 90 MHz. The measured frequency interval of the two microwave signals when the set values are (e) 500 MHz and (f) 1 GHz.

Finally, the SBS-based frequency measurement system is performed to verify the validity of the proposed approach. In this experiment, the frequency interval of two microwave signals is measured using the filter- and FTTM-based frequency measurement system in Fig. 4, and the frequency measurement accuracy of the system is evaluated by comparing the measured frequency interval with the actual interval preset. To break the frequency resolution and accuracy limitation imposed by fast frequency sweep, the SBS gain bandwidth broadening is also introduced in this experiment. In addition, the sweep rate of the probe path is set to 4 GHz/μs.

A two-tone signal with a frequency interval of 500 MHz is chosen as the SUT. Fig. 11 (a) and (b) show the measured time-domain pulses and corresponding time-frequency diagrams when the sweep ranges of the pump wave are set to 0 and 90 MHz. As can be seen, with the broadening of the SBS gain bandwidth through increasing the sweep range of the pump wave from 0 to 90 MHz, the pulse width is greatly reduced from about 63.4 to 13.8 ns. Under certain other conditions, reducing the pulse width will help to improve the frequency resolution [17]. In addition, the reduction of the pulse width is important to find the time represented by the pulse more accurately, resulting in a more accurate frequency identification. Fig. 11(e) shows the measured frequency interval with and without SBS gain bandwidth increase. As can be seen, the measurement error can be greatly reduced from ±40 MHz to ±4 MHz when employing the SBS gain bandwidth increase. Another two-tone signal with a frequency interval of 1 GHz is further measured, with the results shown in Fig. 11 (c) and (d). The pulse width is also greatly reduced in this measurement with the increase of the SBS gain bandwidth. Fig. 11 (f) shows the measured frequency interval with and without SBS gain bandwidth increase in this case. The measured accuracy is also improved by about 10 times from ±40 MHz to ±4 MHz.

Note that, in Fig. 11, both time-domain pulses in a single sweep period and the time-frequency diagram accumulated with time are given. For frequency measurement, a fast sweep in a single period is enough for obtaining the frequency.

*3.5 Comparison*

Table 1 shows the performance comparison of the reported filter- and FTTM-based time and frequency acquisition systems. Most of the previously reported results are demonstrated under a relatively low sweep rate of hundreds or tens of gigahertz per second or even less. Although in these cases, the measurement resolution and accuracy have been demonstrated to be better than 10 and 1 MHz [16, 17], which is very good results. However, in some application scenarios, a much higher sweep rate of the measurement system is highly required. In some works [21, 22], although the sweep rate is much increased, the measurement performance is much decreased compared with the system using a much slower sweep rate. In addition to the one-dimensional frequency measurement systems, the filter- and FTTM-based two-dimensional time and frequency measurement system [25] also has a high requirement on the sweep rate to obtain approximate stationarity in a short time. However, using the conventional methods employing a nature SBS gain spectrum has limited frequency resolution, especially when the sweep rate is very high.

Comparing this work with the previously reported ones, it is clear that, when the sweep rate is on the same order of magnitude, the measurement accuracy of the frequency measurement is improved by around 25 times (this work and [21]), whereas the frequency resolution of the STFT in this work is also doubled or tripled compared with our former work in [25].

Table 1. Performance Comparison of Filter-based FTTM Measurement Systems

| | Error (MHz) | Resolution (MHz) | Sweep rate (GHz/μs) | Filter BW (MHz) | STFT/ MFM |
|---|---|---|---|---|---|
| [13] | 90 | 200 | $1.25 \times 10^{-4}$ | 170 | |
| [14] | 237.3 | 375 | $3.3 \times 10^{-3}$ | 325 | |
| [15] | / | 90 | $5.4 \times 10^{-3}$ | 78.5 | |
| [16] | ±1 | 25 | $5 \times 10^{-5}$ | 20.3 | |
| [17] | ±1 | 10 | $1.6 \times 10^{-5}$ | 10.3 | MFM |
| [18] | ±100 | 250 | /[a] | | |
| [19] | ±0.4 | 1 | $2.5 \times 10^{-6}$ | 0.75 | |
| [20] | ±60 | 60 | 4/22.22 | 60 | |
| [21] | ±100 | 200 | 2.74 | 30 | |
| [22] | / | 60[b] | 1 | 20.3 | STFT |
| | | 300[b] | 4 | | |
| This work | ±4 | / | 4 | 64 | MFM |
| | / | 35[b] | 1 | 38 | STFT |
| | | 105[b] | 4 | 64 | |

[a] The method is based on frequency shifting recirculating delay line.

[b] The value is obtained using a two-tone test, under the corresponding sweep rate and filter bandwidth.

## 4. Conclusion

In summary, for the first time, the accuracy and resolution limitation of filter- and FTTM-based time and frequency acquisition methods in the fast sweep scenario is broken by broadening the filter bandwidth. The principle of the method is comprehensively analyzed and verified by a series of experiments using an SBS-based system. It is found that when the sweep rate of the system is fixed, a suitable filter bandwidth can always be found to minimize the width of the generated pulse signal, thus maximizing the frequency resolution accordingly. When a fast measurement is employed, this phenomenon is more useful because the optimal filter bandwidth is often larger and easier to realize in the optical domain. Microwave frequency measurement and STFT are implemented and compared with the results without SBS gain bandwidth increase. The frequency measurement accuracy of the system is improved by around 25 times compared with the works using a similar sweep speed, while the frequency resolution of the STFT is also much improved compared with our former work. The method proposed in this work is of great significance and guidance to the frequency and time-frequency measurement system based on FTTM and frequency sweeping. It solves the contradiction between the sweep speed and measurement accuracy and resolution, and is of great help to quickly and accurately obtain the frequency and time-frequency information of unknown electromagnetic signals.

**Acknowledgements**

This work was supported by the National Natural Science Foundation of China [grant number 61971193]; the Natural Science Foundation of Shanghai [grant number 20ZR1416100]; the Science and Technology Commission of Shanghai Municipality [grant number 18DZ2270800].